\documentclass[10pt]{article}

\usepackage{hyperref}
\hypersetup{backref,colorlinks=true}
\usepackage{graphicx}

\begin{document}

\title{\textbf{Finding the best proxies for the solar UV irradiance}}

\author{T. Dudok de Wit$^{1)}$, M. Kretzschmar$^{1)}$, J. Lilensten$^{2)}$, T. Woods$^{3)}$}

\date{\small $^{1)}$ LPC2E, CNRS and University of Orl\'eans, Orl\'eans, France \\
$^{2)}$ LPG, CNRS and Joseph Fourier University, Grenoble, France \\
$^{3)}$ LASP, University of Colorado, Boulder, USA}

\maketitle

\begin{abstract}
Solar UV emission has a profound impact on the upper terrestrial atmosphere. Because of instrumental constraints, however, solar proxies often need to be used as substitutes for the solar spectral variability. Finding proxies that properly reproduce specific spectral bands or lines is an ongoing problem. Using daily observations from 2003 to 2008 and a multiscale statistical approach, we test the performances of 9 proxies for the UV solar flux. Their relevance is evaluated at different time-scales and a novel representation allows all quantities to be compared simultaneously. This representation reveals which proxies are most appropriate for different spectral bands and for different time scales.

\bigskip

\noindent
preprint of an article accepted for publication in \textit{Geophysical Research Letters} (2009) \\
DOI:10.1029/2009GL037825

\bigskip
\end{abstract}


\section{Why are solar proxies needed ?}

The solar irradiance in wavelengths shortward of 300 nm is a key parameter for the specification of the upper terrestrial atmosphere \cite{floyd02}. Variations are observed on time-scales ranging from seconds to years and can impact radio wave propagation, satellite orbits through increased air-drag but also global Earth climate. Unfortunately, there has been no long-term and continuous measurement of the full solar UV spectrum until Feb. 2002, when the TIMED satellite started operating. Even today, the continuous measurement of solar irradiance with sufficient temporal resolution and radiometric accuracy remains a major instrumental challenge  \cite{woods05}. A important issue is the identification of proper substitutes (i.e. proxies) of the solar UV flux for upper atmospheric modeling \cite{lilensten08}. 

Here, we test the performance of nine proxies for various UV spectral bands that encompass emissions coming from the solar corona down to the photosphere. The variability should therefore be strongly wavelength dependent, even though different solar layers are coupled. The solar UV radiation mostly affects the terrestrial atmosphere through photoionization and photochemistry, which are again wavelength-dependent processes \cite{floyd02,lilensten08}. The radiation in the MUV range (200--300 nm) mostly affects the stratospheric O$_3$ concentration; the FUV range (122--200 nm) affects the upper mesospheric O$_2$ excitation production and the lower thermospheric O$_2$ dissociation; the EUV range (10--120 nm) affects the thermospheric O, O$_2$ and N$_2$ ionization and excitation productions. Other effects include the impact of the intense H I Lyman-$\alpha$ line at 121.57 nm on nitric oxides, which are important for climatological considerations. The altitude of strongest absorption is shown in Fig. 1, together with the average spectral irradiance. No single proxy can reproduce the solar variability over the whole UV spectrum. Our prime objective therefore is to compare the measured irradiance in these bands to various  proxies that are measured by independent means, partly from ground instruments. Several of these proxies have the advantage of not suffering from instrument degradation and so are easier to maintain in the long.

\begin{figure}
\noindent\includegraphics[width=22pc]{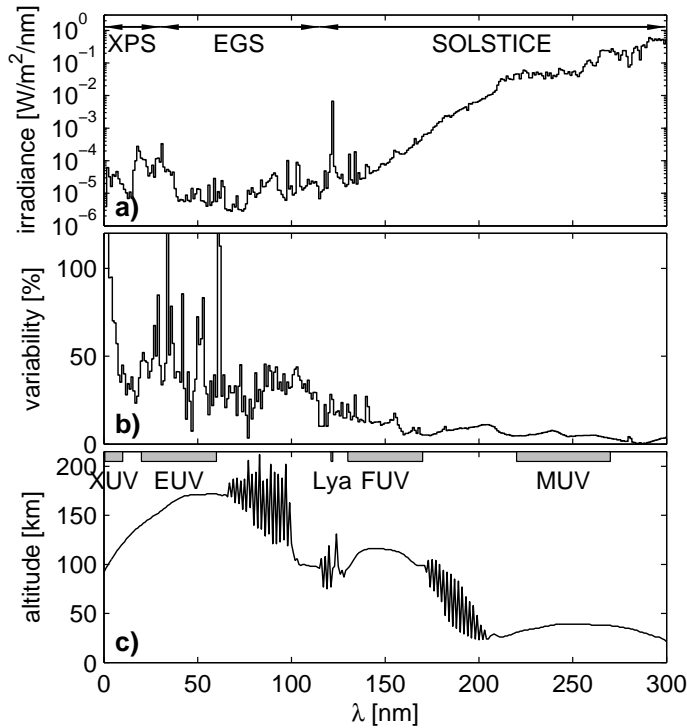}
\caption{a) Average solar irradiance for the considered time interval; b) Variability, defined here as $100 \times$(maximum-minimum)/average irradiance over the period Aug. 2003 -- Feb. 2008; c) Altitude of unit optical depth, an indicator of the altitude of maximum atmospheric absorption. The spectral bands refer to the naming conventions used in Sec.~2.}
\end{figure}

Many authors have already evaluated the relevance of solar proxies for upper atmosphere specification. Physical models of the solar irradiance (e.g. \cite{fontenla07}) are not yet accurate enough below 400 nm, so most studies are based on experimental data, using statistical analyses or by comparing specific events  \cite{parker98,kane02c,floyd05}. Some authors have tested upper atmospheric models with different solar inputs \cite{thuillier01}. Various empirical models use such proxies to deliver solar UV spectra  \cite{tobiska00,richards06}.

Our approach is statistical as well. We use daily spectra from the TIMED and SORCE missions, which provide uninterrupted coverage of the full UV spectral range. In contrast to previous studies, however, we first decompose the different quantities into several time-scales before comparing them. Indeed, since different time-scales (e.g. solar rotation period and solar cycle) capture different physical processes, a multiscale approach is needed. The second novelty is a graphical representation, similar to the one used by \cite{ddw05}, which, for the first time gives a global picture of how all proxies and spectral bands are related to each others, without having to go through the tedious visual comparison of scatter plots or correlation coefficients.


\section{The data and the method}

We consider daily measurements, covering the declining phase of the solar cycle, from from Aug. 2, 2003 until February 1, 2008. Flares are not included here because most instruments don't properly resolve them. The Solar flare \cite{tobiska05} and FISM \cite{chamberlin08} models have been specifically designed to model the short term spectral variability during flares by incorporating high cadence soft X-ray measurements.

Our spectral measurements are a composite of three different instruments. The 1--27 nm and the 121--122 nm ranges are covered by the XPS photometers onboard SORCE \cite{woods05b}. The data (version 9) in that range are not purely observational since the CHIANTI model is used to compute the irradiance in 1 nm bins. The 27--115 nm range is covered by the EGS spectrograph onboard TIMED \cite{woods05}, whose spectral resolution is 0.4 nm (version 9, the data are rebinned to 1 nm). The remaining wavelengths are covered by the SOLSTICE spectrometers onboard SORCE \cite{rottman06b} with a resolution of 1 nm (version 16).

In the following, we concentrate on 5 spectral bands that are considered to be important for aeronomy \cite{lilensten08,tobiska08}, see Fig. 1. We shall call them XUV (0.5--10 nm), EUV (20--60 nm), Lyman-$\alpha$ (121--122 nm), FUV (130--170 nm) and MUV (220--270 nm). Note that these definitions differ from the ISO 21348:200 standard that was used in Sec.~1. The proxies are:
\begin{enumerate}
\item \textit{ISN}, the international sunspot number (from SIDC, Brussels), which is not really a UV proxy but remains the most widely used gauge of solar activity.
\item \textit{f10.7} is the radio flux at 10.7 cm (from Penticton Observatory, Canada). This proxy is widely used as a solar input to ionosphere/thermosphere models, partly because it can be conveniently measured from ground.
\item \textit{MgII} is the core-to-wing ratio of the Mg II line at 280 nm (from SORCE/SOLSTICE, version 9). This index probes the high chromosphere and is often advocated for the FUV \cite{heath86}.
\item \textit{CaK} is the normalized intensity of the Ca II K-line at 393 nm (from National Solar Observatory  at Sacramento Peak). This line originates at nearly the same altitude as the Mg II line and has also been advocated for the FUV \cite{lean82}.
\item \textit{MPSI} is the magnetic plage strength index (from the Mt. Wilson 150-Foot Solar Tower), which quantifies the relative fraction of the solar surface that is covered by mild magnetic fields ($10 < |\mathbf{B}| < 100$ Gauss). By definition, the MPSI index is a proxy for plages and faculae \cite{parker98}.
\item \textit{MWSI} is the Mount Wilson sunspot index, defined as the MPSI, but for intense magnetic fields ($|\mathbf{B}| > 100$ Gauss). The MWSI is a proxy for active regions.
\item \textit{s10.7} is computed by \cite{tobiska08} out of the integrated 26-34 nm emission from the SEM radiometer onboard SoHO, and rescaled to the f10.7 index after a trend correction (version 3.9a). This proxy is dominated by the emission from the chromospheric and transition region  {He II} line at 30.4 nm. 
\item \textit{Lyman-$\alpha$ channel} (ch-L) is the expected output of a photodiode from the LYRA radiometer \cite{hochedez06} onboard the PROBA2 satellite, whose launch is scheduled for the end of 2009. This channel integrates emissions in the 110-210 nm band, with a peak around 125 nm. We reconstructed the signal from SORCE/SOLSTICE data.
\item \textit{Herzberg channel} (ch-H) is the expected output of a photodiode from LYRA in the Herzberg band, here between 195--220 nm. Both the Herzberg and the Lyman-$\alpha$ channels are relevant inputs for upper atmospheric models.

\end{enumerate}

\medskip

Note that the last three quantities are actually derived from UV irradiance measurements and so do not really qualify as proxies. Together with the MgII index, they must be measured from space.

To compare different time-scales, we first decompose each quantity into different scales using the \textit{\`a trous} wavelet decomposition \cite{mallat98}. The decomposition imposes scales that are centered on 3, 6, 12, 24, \ldots, 768 days. The components associated with scales of 24 and 96 days are illustrated in Fig. 2. The first scale captures solar rotation effects and the second one the long-term evolution of active regions. The most conspicuous result in Fig. 2 is the remarkable coherence of all quantities, which justifies \textit{a posteriori} the widespread use of proxies for the solar UV flux.

All quantities are affected by various types of noise. The signal-to-noise ratio of the irradiance degrades as the flux drops with decreasing solar cycle and detector degradation increases. Data gaps in the CaK (with 22\% coverage and a median gap duration of 3 days), MPSI and MWSI (75 \% coverage, 1 day median) indices were filled by multivariate interpolation \cite{kondrashov06}. Gap filling affects the variability on the shortest time-scales (up to about 6 days) by enhancing the correlation with other proxies. The smallest scales require more careful analysis and are therefore discarded here.

\begin{figure}
\noindent\includegraphics[width=22pc]{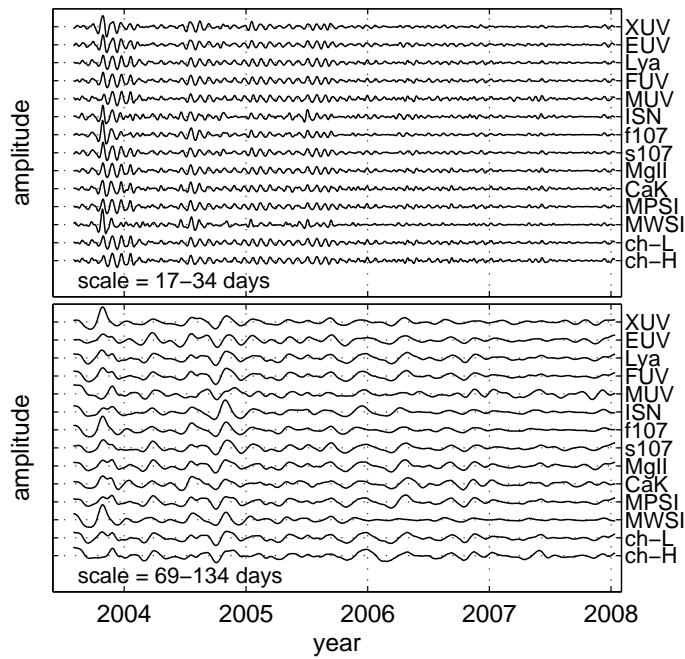}
\caption{Decomposition of the 5 spectral bands and the 9 proxies into two different scales: solar rotation period (top) and an average time-scale of 96 days (bottom). Amplitudes are arbitrary. The legends are given in the text.}
\end{figure}


\section{The proxy-to-spectral band relation}

Since all quantities are strongly correlated, we use the Pearson correlation coefficient $r_{xy} = s_{xy}/(s_x s_y)$ (where $s_{xy}$ is the sample covariance, and $s_x^2$ is the sample variance) as a gauge of linear correspondence between $x$ and $y$. Unfortunately, this quantity can only be computed pairwise, leaving us with a large table of values to decipher. Better insight can be gained by turning these values into a graph, using the multidimensional scaling technique \cite{borg97}. The idea consists in representing all irradiances and proxies on a 2D correspondence map in such a way that their mutual distance equals their dissimilarity, defined here as $d = 1-|r|$. Such a map allows the proxy-to-spectral band relation to be investigated in detail. Two close variants of this approach have already been used by \cite{ddw05,ddw08}.

In principle, we would need a 13-dimensional plot to represent all quantities while accounting for their mutual distances exactly \cite{borg97}. However, because of the strong coherency of the spectral variability, a 2D (and sometimes even a 1D) approximation already captures the salient statistical features. We verified that a third dimension does not bring new insight here.

Figure 3 is the central result of this study, as it shows the correspondence between proxies and spectral bands for three typical scales. What matters in these plots is the distance between each pair, and not the absolute positions or the axes, which do not carry any direct meaning. The closer two quantities are, the more correlated they are. The correspondence map therefore provides both quantitative (the distance is related to $r$) and qualitative information on how all quantities are related to each other. The top panel covers time-scales of typically half a solar rotation, the middle panel one solar rotation and the bottom panel several rotations.  It comes as no surprise that quantities with strong physical connections, such as the Lyman-$\alpha$ band and the Lyman-$\alpha$ channel, always stick together. 


\begin{figure}
\noindent\includegraphics[width=22pc]{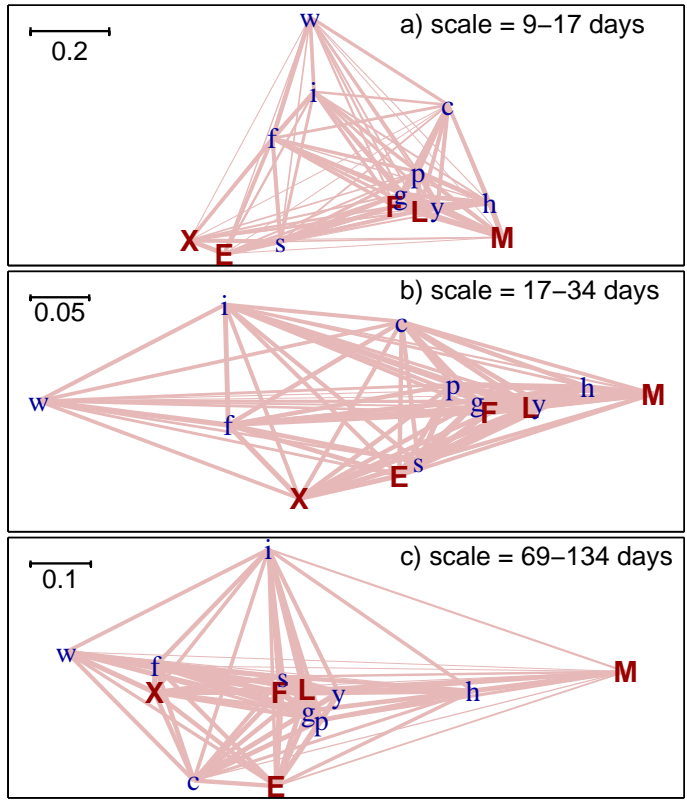}
\caption{Correspondence maps for three characteristic scales. The distance between each pair of points corresponds to $1-|r|$ (see text) and the line thickness is proportional to $r$. Capitals designate our 5 spectral bands: (X)UV, (E)UV, H I (L)yman-$\alpha$, (F)UV and (M)UV. The other characters correspond to proxies: (i)sn, (f)10.7, (s)10.7, M(g)II, (c)aK, M(p)SI, M(w)SI, L(y)man-$\alpha$ channel, (h)erzberg channel.}
\end{figure}


Interestingly, all spectral bands are roughly distributed along a line, revealing a gradual transition from optically thin and energetic emissions on the left to optically thick ones on the right. For time-scales below one solar rotation (upper plot), this ordering manifests itself in the center-to-limb variation, with brightenings in the XUV and EUV bands, and darkenings in the FUV and MUV bands \cite{donnelly90}. Our maps show that the same gradual transition persists for longer time-scales. The relative location of proxies with respect to this alignment changes substantially with time-scale, which confirms and highlights the importance of a multiscale approach.

The bottom panel focuses on time-scales that encompass the decay time of active regions. Similar distributions are observed for larger scales but the solar cycle time scale could not be addressed by lack of data. Some correlations are difficult to explain, such as the close location of the MgII index and the EUV band, already noted by \cite{judge02}. Equally surprising is the difference between the MgII and CaK indices, which probe nearly the same solar altitude.


\section{Conclusions and Recommendations}

What are the best proxies for reconstructing specific spectral bands ? Figure 3 precisely tells us that the answer very much depends on the time-scale of interest. Several general conclusions can nevertheless be drawn.
\begin{enumerate}
\item Proxies that are derived from real irradiance data do indeed match their corresponding spectral band quite well. The correlation between the LYRA Lyman-$\alpha$ channel and the Lyman-$\alpha$ band, for example, always exceeds 0.95.
\item The MgII index shows the best global performance since it is always located close to the center of the cloud of points. It is particularly well suited for the FUV band. The MPSI index is a backup solution, but is not measured continuously.
\item Apart from these, however, no single spectral band can be properly reconstructed at all scales from one single proxy.    
\item There is no good (non irradiance-derived) proxy for the XUV and EUV bands. The f10.7 index would be the least bad solution. 
\item None of our proxies properly fits the MUV band, for which a better gauge of photospheric emissions is needed.
\end{enumerate}

\medskip

The correspondence map can also be used to search for new proxies. Indeed, when three quantities are closely spaced and aligned, then the one in the middle be approximated by taking a linear combination of the two others. Two spectral bands whose reconstruction can be improved that way are the FUV and the Lyman-$\alpha$ bands. Reciprocally, remotely located quantities are hardest to reconstruct. This is particularly evident for the MUV and the XUV bands. For the same reason, the sunspot number and the MWSI  should not be used for any reconstruction as they capture variations that are not reproduced by the irradiance.

We are currently developing this multiscale approach for more accurate reconstructions of the previously considered UV bands. The EUV band, for example, can be reconstructed as a sum of contributions from different proxies, decomposed into different scales and with different weights. Our tests show that the residual error between the observed and modeled flux can be reduced that way by a factor of 2 to 4 as compared to models that do not consider multiple scales. These results will be detailed in a forthcoming publication. A second improvement consists in incorporating the effect of a convolution with time, as studied by \cite{preminger07}. An aspect that matters for operational models is the availability of the proxies, since their long-term measurement is not always guaranteed.

\subsection*{Acknowledgments}
We thank two anonymous referees for their helpful comments. The following institutes are acknowledged for providing the data: Solar Influences Data Center (Brussels), Laboratory for Atmospheric and Space Physics (Boulder), National Geophysical Data Center (NOAA), National Solar Observatory at Sacramento Peak (data produced cooperatively by NSF/NOAO, NASA/GSFC and NOAA/SEC) and Mount Wilson Observatory (operated by UCLA, with funding from NASA, ONR  and NSF, under agreement with the Mt. Wilson Institute). Thanks are also due to K. Tobiska (Space Environment Technologies) for providing the s10.7 index and I. Dammasch (SIDC) for providing the passbands of LYRA.  This study received funding from the French PNST programme and from the European Community's Seventh Framework Programme (FP7/2007-2013) under the grant agreement nr. 218816 (SOTERIA project, www.soteria-space.eu).

\bibliographystyle{abbrv}

\end{document}